\documentclass[twocolumn,superscriptaddress]{revtex4-2}
\usepackage[utf8]{inputenc}
\usepackage{graphicx,braket,amsmath,amssymb,nicefrac,braket,xcolor}

\global\long\def\ket#1{\left|#1\right\rangle }%

\global\long\def\bra#1{\left\langle #1\right|}%

\global\long\def\braket#1#2{\left\langle #1\left|#2\right.\right\rangle }%

\begin{document}

\title{Probing The Unitarity of Quantum Evolution Through Periodic Driving}

\author{Alaina M. Green}
\affiliation{Joint Quantum Institute and Department of Physics, University of Maryland, College Park, MD 20742, USA}
\author{Tanmoy Pandit}
\affiliation{Fritz Haber Research Center for Molecular Dynamics, Hebrew University of Jerusalem, Jerusalem 9190401, Israel}
\author{C. Huerta Alderete}
\affiliation{Joint Quantum Institute and Department of Physics, University of Maryland, College Park, MD 20742, USA}
\author{Norbert M. Linke}
\affiliation{Joint Quantum Institute and Department of Physics, University of Maryland, College Park, MD 20742, USA}
\affiliation{Duke Quantum Center and Department of Physics,
Duke University, Durham, North Carolina 27708, USA}
\author{Raam Uzdin}
\affiliation{Fritz Haber Research Center for Molecular Dynamics, Hebrew University of Jerusalem, Jerusalem 9190401, Israel}

\date{\today}

\begin{abstract}

As quantum computers and simulators begin to produce results that cannot be verified classically, it becomes imperative to develop a variety of tools to detect and diagnose experimental errors on these devices. While state or process tomography is a natural way to characterize sources of experimental error, the intense measurement requirements make these strategies infeasible in all but the smallest of quantum systems. In this work, we formulate signatures of unitary evolution based on specific properties of periodically driven quantum systems. The absence of these signatures indicates a break either in the unitarity or periodicity condition on the evolution. We experimentally detect incoherent error on a trapped-ion quantum computer using these signatures. Our method is based on repeated measurements of a single observable, making this a low-cost evaluation of error with measurement requirements that scales according to the character of the dynamics, rather than the system size.

\end{abstract}

\maketitle

\section{Introduction}

The utility of state-of-the-art quantum computers (QCs) is limited by experimental noise which degrades their accuracy. While methods of quantum error correction are rapidly advancing towards the ability to correct for experimental noise~\cite{sho95,kni97,chi04,ree12,cam20,egan21}, reducing the errors remains a high priority and understanding the characteristics of the noise is indispensable in reducing or eliminating its effect on the computational result~\cite{bru19,kra19,fao04,kot13,sun19}.

While the physical errors affecting different types of quantum computers are sometimes understood in a general sense (such as electric field noise in trapped-ion QCs~\cite{bru19} or Johnson noise in superconducting QCs~\cite{kra19}), the specific errors limiting the fidelity of any particular quantum computer are unique and therefore more challenging to determine. 

The noise can be fundamental -- arising from intrinsic properties of the chosen qubit platform -- or technical -- arising from  imperfect execution of the quantum control on which the computing relies. Additionally, the noise can be coherent or incoherent, which can be a critical distinction for the feasibility of error correction and mitigation protocols~\cite{leu97,caf14,pur20,cal96,sho95,ste96,kni97,tem17,li17,hug21,cai21,low21}.

Basic knowledge of the physics governing the operation of the QC often suggests the character of a particular source of physical noise across the categories mentioned above -- intrinsic vs. technical and coherent vs. incoherent. As an example, consider the trapped ion quantum computer (TIQC) in Ref.~\cite{deb16}. It uses laser-driven electronic transitions to orchestrate entanglement through phonons in a trapped ion chain~\cite{sor99,sor00}. These entangling operations are fundamentally limited by spontaneous photon scattering off electronic spin states outside the qubit manifold, an incoherent noise source which occurs during the quantum dynamics~\cite{oze07,bal16}. However, the lasers used to drive this entanglement can also lead to technical, coherent errors in quantum dynamics if the laser intensity unexpectedly changes, leading to miscalibration. Therefore, an operational test that distinguishes coherent from incoherent errors could discern if the TIQC is limited by fundamental or technical noise, guiding possible improvement.

There are already a number of distinct strategies for characterizing error on quantum devices~\cite{eis20,kni08,blu17,erh19,pro22}. Quantum gate set tomography can be used as a sensitive, system-wide diagnostic of error~\cite{blu17,nie21}. However, it requires a non-scalable number of gate operations and measurements~\cite{das11}. In contrast, randomized benchmarking (RB) and its extensions approximate the average error in a sequence of quantum gates by repeatedly applying a unitary followed by its inverse and measuring the deviation of this action from the identity~\cite{kni08,mag11,cro16,pro19,xue19,hel19,pro21,erh19}. However, the randomization partially obscures the character of the noise, requiring careful interpretation of which model best fits the data to understand if the noise is coherent or not~\cite{wal15,she16}. Additionally, RB and its extensions cannot be straightforwardly extended beyond universal quantum computers, such as to analog simulators~\cite{sha21}.

In this paper we demonstrate a scheme capable of distinguishing one subset of possible errors from another using a number of measurements, which does not necessarily increase with the system size. Specifically, this scheme is sensitive only to incoherent error or coherent errors arising from a drift in control parameters. Further, the ability of this scheme to detect these two particular errors can be increased by taking more measurements.

Previously, we reported that single basis measurements of a quantum state taken at multiple points during periodic evolution can be used to detect incoherent errors~\cite{pan22}. Now we explicitly derive an expression composed of these time-series measurements of periodic operator evolution (POE) which exhibits unique behavior under two assumptions: that the evolution is unitary and periodic. Hence, any deviation from this behavior indicates the presence of either uncontrolled parameter changes leading to non-periodicity or an incoherent noise source, hereby collectively referred to as POE-sensitive errors. For this demonstration, the periodic evolution is a short series of quantum gates applied repeatedly, as shown in Fig.~\ref{fig:LongEvolve}a. However, we emphasize that our scheme can work for other forms of periodic unitary driving, such as that performed in analog quantum simulators, for which methods such as RB do not apply.

The remainder of this paper is organized as follows. In Sec.~\ref{sec:POEMeasurementsandInequality} we outline the procedure for obtaining the POE measurements based on the probability that the system is found in its initial state after periodic driving, demonstrating how they can be analyzed to detect certain errors in the form of an inequality. We present our main result in Sec.~\ref{sec:ExpDecay}, a constraint on the behavior of the POE measurements in the form of an exponential scaling law which has the advantage of requiring less evolution than the inequality to detect errors in practice, along with the further benefit of being highly resistant to state preparation and measurement errors.

We use the exponential scaling law from Sec.~\ref{sec:ExpDecay} to demonstrate the detection of intentionally added incoherent error in Sec.~\ref{sec:expMain}. Before concluding in Sec.~\ref{sec:Conclusions}, we provide extensions to our main results in Sec.~\ref{sec:Extensions}, including a scheme based on the probability that the system has transitioned to a different state, and error characterization for a quantum system based on measurements only of a sub-system.

\section{Inequalities from POE Measurements\label{sec:POEMeasurementsandInequality}}

\begin{figure}
	\includegraphics[width=0.97\columnwidth]{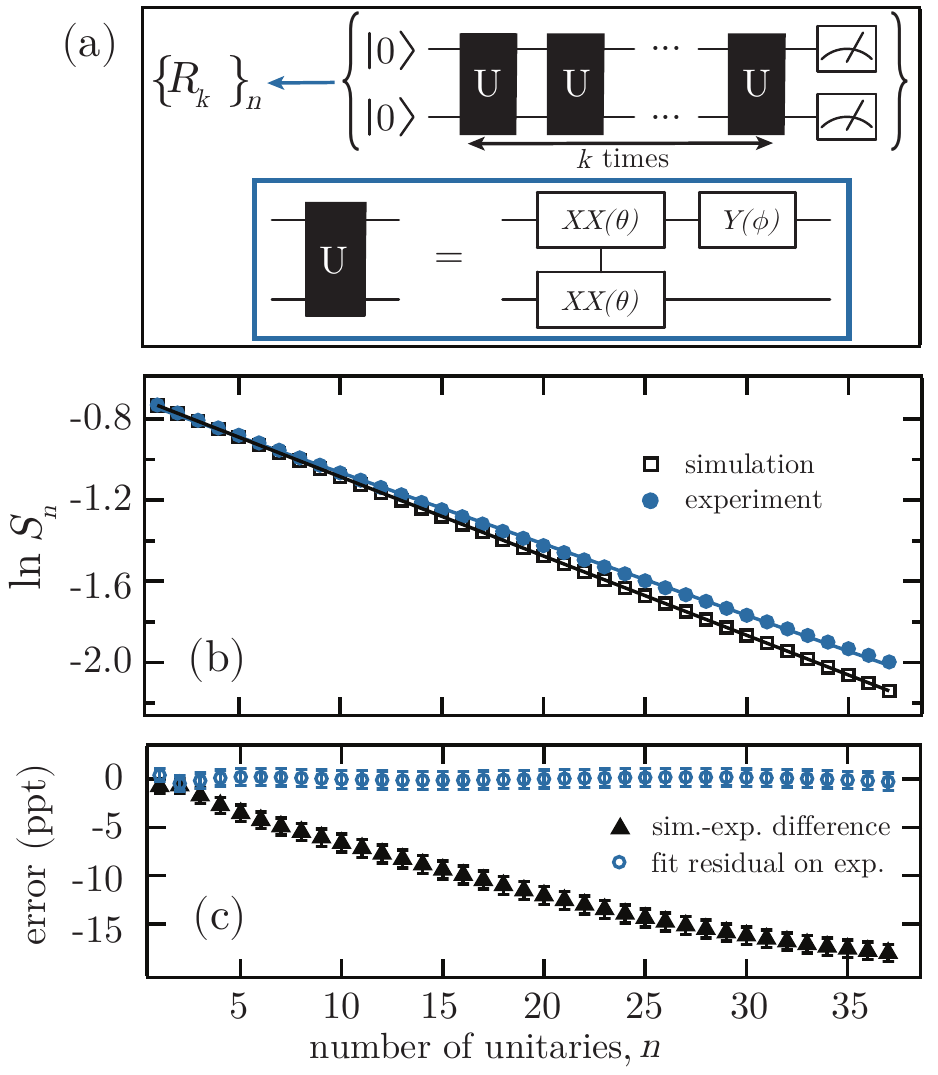}
    \caption{(a) Sketch of the experimental procedure of the POE measurement used to test the exponential scaling law of $S_n$. For each $R_k$ we measure after $n$ applications of $U$. Also shown is the arbitrarily selected unitary in the form of two quantum gates: one entangling gate and one single-qubit rotation, described in Appendix~\ref{sec:ExpDet}. (b)~Experimental (closed blue circles) and simulated (open black squares) values of $\ln S_n$ for increasing numbers of $n$. Exponential fits to the data are shown in solid lines. (c) A comparison of two ways to quantify errors on $S_n$ in parts per thousand. The difference between the simulated value of $S_n$ and the experimental value of $S_n$ (closed black triangles) is significantly larger in magnitude than the residual of the exponential fit (open blue circles), indicating that the errors are dominantly coherent and periodic, i.e. POE-insensitive. Note that while the fit residual has some structure, it is consistent with zero given our statistical uncertainty.}
	\label{fig:LongEvolve}
\end{figure}

The POE measurements upon which our constraints are built are pictured in Fig.~\ref{fig:LongEvolve}a. Initialized in an arbitrary state, a quantum system of arbitrary size is evolved under some unitary $U$ and measured in a single basis to measure the recurrence probability, the likelihood that the system is found in its initial state, also known as the survival probability. This entire process is repeated multiple times, each iteration adding one more application of $U$. Working in Liouville space, the recurrence probability is denoted $R_{n}=\braket{\rho_{0}}{\rho_k}$ where $\rho_0$ is the initial density matrix and $\rho_{k}$ is the density matrix after the $k^{\mathrm{th}}$ application of $U$~\cite{gya20}. 

Our first constraint from POE measurements is based on the definition of positive operators: an operator $O$ is positive if and only if $\bra{\psi}O\ket{\psi}\geq 0$ for any $\ket{\psi}$. We will formulate $O$ such that POE measurements are constrained by this inequality. We begin by constructing a positive operator based on the positive function, $f^n=\left(\frac{1}{2}-\frac{1}{2}\cos{x}\right)^n$. By rewriting $\cos(x)$ in exponential form and recasting $x$ as a $\frac{Ht}{\hbar}$, we obtain an explicitly positive operator, $F$, defined through the evolution of some effective Hamiltonian, $H$, which generates the unitary $U=e^{-\frac{i}{\hbar}Ht}$:

\begin{equation}
    F^n=\left(\frac{1}{2}-\frac{U+U^\dagger}{4}\right)^n
    \label{eq:FGton}
\end{equation}

Because $U$ is unitary, $F^n$ can be expanded into the following form (see Appendix~\ref{sec:coefderiv} for details):

\begin{equation}
    F^n=\frac{1}{2^{2n}}\binom{2n}{n}+\frac{1}{2^{2n}}\sum_{k=1}^{n} (-1)^k \binom{2n}{n-k} \left( U^k + U^{\dagger k} \right).
    \label{eq:NegSum}
\end{equation}
   
\noindent Measuring $F^n$ relative to the initial state provides an inequality composed of POE measurements, which we call $S_n$:

\begin{equation}
    S_n \equiv \bra{\rho_0} F^n \ket{\rho_0} \geq 0.
    \label{eq:SnIneq}
\end{equation}

Combining Eqs.~\ref{eq:NegSum} and~\ref{eq:SnIneq} we obtain:

\begin{multline}
    S_n = \frac{1}{2^{2n}}\binom{2n}{n} \braket{\rho_0}{\rho_0} +\\
    \frac{1}{2^{2n}}\sum_{k=1}^{n} (-1)^k \binom{2n}{n-k} [ \bra{\rho_0} U \ket{\rho_0} + \bra{\rho_0} U^\dagger \ket{\rho_0}] \geq 0
    \label{eq:finalinequal}
\end{multline}

\noindent To finalize the expression of $S_n$ in terms of the recurrence probability, we note that because the evolution is periodic, $\bra{\rho_0}U^k \ket{\rho_0}=\bra{\rho_0}U^{\dagger k} \ket{\rho_0}=\braket{\rho_0}{\rho_k} = R_k$ for all $k$. This results in:

\begin{equation}
    S_n=\frac{1}{2^{2n}}\binom{2n}{n}R_0 + \frac{1}{2^{2n-1}}\sum_{k=1}^{n} (-1)^k \binom{2n}{n-k} R_k \geq 0.
    \label{eq:inequality}
\end{equation}

\noindent Hence, by making a single-basis measurement after $n$ intervals in periodic evolution, we can build an inequality, which holds if the evolution is unitary and periodic. By constructing Eq.~\ref{eq:inequality} from positive operators, we not only provide an explicit analytical expression for the inequalities implicitly derived in~\cite{pan22}, but we also enable the following spectral analysis, which leads to the exponential decay of $S_n$ in the absence of POE-sensitive errors.

\section{Exponential Decay of POE Measurements\label{sec:ExpDecay}}

In this section we show that $S_n$ decays exponentially under ideal conditions and outline how this fact can be used to detect incoherent error or uncontrolled parameter changes.

\subsection{Proof by Spectral Analysis}

$F^n$ can be written in the eigenbasis of $F$:

\begin{equation}
    \label{eq:EigSum}
    F^n=\sum^{N^2}_{j=1} \lambda_{j}^{n} \ket{j}\bra{j}
\end{equation}

\noindent where $0 \leq \lambda_{j} \leq 1$ are the eigenvalues of $F$ and $N$ is the Hilbert space dimension. The sum is over $N^2$ because the $N\times N$ density matrix was flattened into $N^2$ density vector $\ket{\rho}$. Using this expression, $S_n$ becomes

\begin{equation}
    S_n=\bra{\rho_0} \sum^{N^2}_{j=1}\lambda_{j}^n \ket{j}\braket{j}{\rho_0}=\sum^{N^2}_{j=1}\lambda_{j}^n|\braket{\rho_0}{j} |^2. \label{eq:EigSumRho}
\end{equation}

\noindent In the limit of large $n$, the largest $\lambda_j$ dominates the sum such that

\begin{equation}
    S_n\approx \lambda_{j_{\mathrm{max}}}^{n} |\braket{\rho_0}{j_{\mathrm{max}}} |^2 \text{    for    } n \gg 1,
    \label{eq:largen}
\end{equation}

\noindent In Appendix~\ref{sec:limit} we show that the value of $n$ for which the scaling law is valid depends on the distribution of the eigenvalue spectrum, and not on the quantum system size directly. Finally, the exponential decay can be easily seen by taking the logarithm to find the dependence on $n$:

\begin{equation}
    \label{eq:scaling}
    \ln \left(S_n\right)\approx 2\ln |\braket{\rho_0}{j_{\mathrm{max}}} | -n\ln{\frac{1}{\lambda_{j_{\mathrm{max}}}}} \text{    for    } n \gg 1.
\end{equation}

\noindent Thus, $S_n$ decays exponentially with increasing $n$ if the underlying assumptions of this derivation are valid. As this signature relies on the scaling instead of the initial condition, it is also robust to errors in state preparation and measurement (SPAM) as is shown in Appendix~\ref{sec:SPAM}.

Finally, we note that similar reasoning can be used to show more generally that $S_n$ is monotonically decreasing and concave for all $n$, another constraint which can be used to detect violations of the unitarity and periodicity assumptions (see Appendix~\ref{sec:monotonic}). 

\subsection{\label{sec:ExpDecayExp} Experimental Demonstration}

We experimentally confirm the exponential scaling law by measuring $S_n$ on a programmable TIQC running a simple periodic circuit. The experimental procedure is pictured in Fig.~\ref{fig:LongEvolve}a, showing that two qubits are initialized in $\ket{00}$ and evolve under repeated applications of the unitary evolution detailed in the blue box. For a definition of the quantum gates used to construct this unitary, see Appendix~\ref{sec:ExpDet}. After each successively longer string of gates is implemented, the state of all qubits is measured in one basis and recorded to get the POE set of recurrence probabilities $\{ R_k \}_n$, which we use to calculate $S_n$. In Fig.~\ref{fig:LongEvolve}b, we plot $\ln S_n$ in order to demonstrate the exponential decay of $S_n$ as a straight line. Fig.~\ref{fig:LongEvolve}c demonstrates the relative magnitude of POE-sensitive errors and non-POE-sensitive errors. The deviation from the correct value of $S_n$, derived from classical simulation of the circuit, is affected by both POE-sensitive and non-POE-sensitive errors while the residual from the exponential fit is only affected by POE-sensitive errors. The difference between the experimental and simulated results being significantly larger than the fit residual indicates that non-POE-sensitive errors are dominant in this case and thus stem either from errors that do not break the periodicity or unitarity of the evolution, for example, miscalibration.

\begin{figure*}
    \centering
	\includegraphics[width=0.97\textwidth]{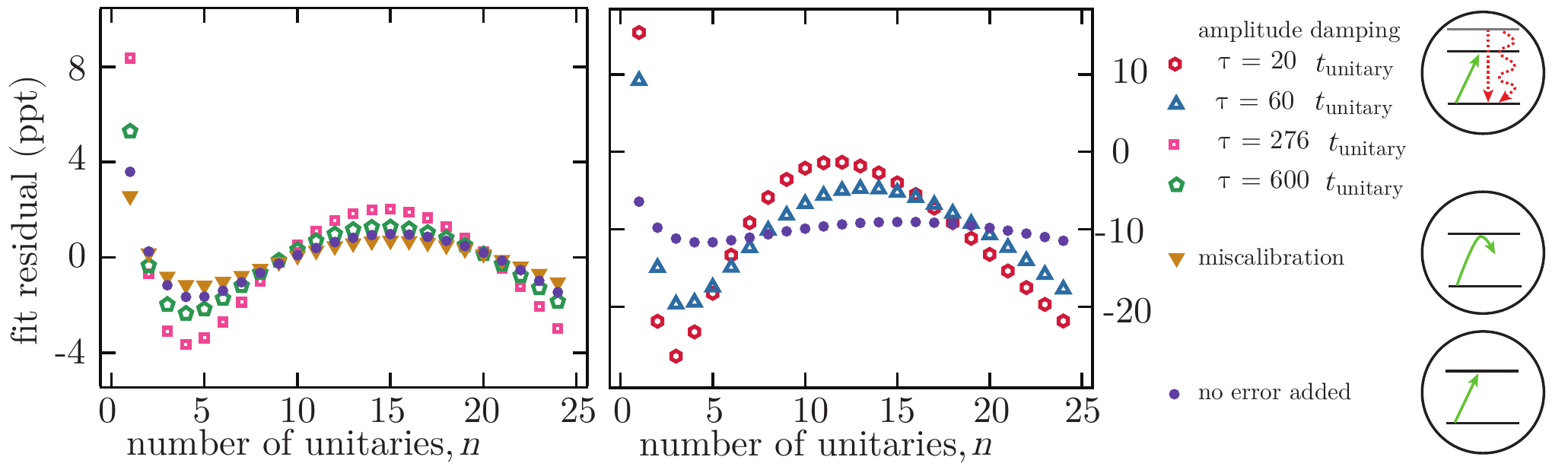}
    \caption{Deviation from exponential decay of $S_n$ under different conditions of added errors as measured by the residual of exponential fits. The types of errors are shown schematically in circular panes. The data, given in parts per thousand, is shown on two different scales for clarity. The experimental procedure and test unitary is the same as in Figure~1a. For comparison, results with no intentionally added noise are shown in closed purple circles. Closed yellow triangles represents a systematic error in calibration of the entangling gate involved in the circuit, corresponding to a consistent over-rotation by 10\% of the angle $\theta$. The open symbols data series correspond to different degrees of added incoherent error in the form of amplitude damping, implemented by optical pumping. The degree of amplitude damping is measured through the lifetime of the $\ket{1}$ state given in terms of $t_{\mathrm{unitary}}$, the time required to implement one unitary. Statistical error bars are similar to those in Fig.~1, about 2 ppt, and are excluded here for visual clarity.}
	\label{fig:DifferentNoises}
\end{figure*}

\section{Incoherent Error Detection From Recurrence Probability\label{sec:expMain}}

The results in Fig.~\ref{fig:LongEvolve} demonstrate that no POE-sensitive error is visible on the scale of $\sim 35$ applications of the unitary for our TIQC. In order to determine the efficacy of this technique, we repeat the experiment shown in Fig.~\ref{fig:LongEvolve}a. in the presence of two types of intentionally added noise of different degrees. The first is the addition of amplitude damping. During the dynamics only, we incoherently drive some portion of the population from $\ket{1}$ to $\ket{0}$ for all qubits (see Appendix~\ref{sec:ExpDet}). This incoherent error breaks the assumptions of the POE scheme and $S_n$ should therefore exhibit behavior which deviates from exponential decay. In Fig.~\ref{fig:DifferentNoises} the results of these experiments are shown as open symbols, with each unique symbol corresponding to a different level of amplitude damping quantified by the lifetime of the state $\ket{1}$. As amplitude damping increases, the value of $S_n$ decays increasingly non-exponentially, indicated by the larger residuals from the exponential fit to $S_n$. 

The second added error is a coherent error occurring at the stage of calibration. We intentionally miscalibrate our entangling gate such that the angle of rotation on the entangling $XX$ gate is $\theta+\delta \theta$ with $\delta \theta$ being a fixed offset which does not drift in time. Our detection scheme is insensitive to this source of error and thus we expect $S_n$ to decay exponentially. The result of this measurement is shown in Fig.~\ref{fig:DifferentNoises} as closed yellow triangles. As expected from our theoretical framework, $S_n$ exhibits nearly exponential decay, similar to the case in which no error is intentionally added, shown in closed purple circles. For reference, the deviation from exponential decay for the purple circles is similar to that shown in blue circles in Fig.~\ref{fig:LongEvolve}b and c, which is known to be a small portion of the total error (see Sec.~\ref{sec:ExpDecayExp}). In comparing the data taken with and without added errors, one is able to see that the data sets with added incoherent error have a greater degree of deviation from exponential behavior than those in which we did not add incoherent errors. The deviation in the case with no added error is likely due to incoherent error which is unavoidable in our apparatus, such as a laser coherence time of 300 ms, which corresponds to the execution time of about 1200 unitaries.

While in this section we have focused on the case in which incoherent noise is the dominant POE-violating error, we note that for other systems, a break in the periodicity condition could be more likely. We also note that incoherent error can be detected through a violation of the inequality in Eq.~\ref{eq:finalinequal}. However, it is not observed experimentally in this paper, highlighting that the exponential scaling law can be experimentally more sensitive.

\section{Extensions to Other Observables\label{sec:Extensions}}

Thus far we have applied our constraints to survival probability measurements on pure initial states. In this section, we consider whether or not our constraints are unique to this observable.

\subsection{Sub-system Probes of Unitarity}

In the previous sections, we have only applied our constraints to pure states of the entire quantum system. In this section, we demonstrate that POE measurements of a sub-system can be used to detect POE-sensitive errors within the whole system provided that the qubits outside the sub-system are in a fully mixed state.

We consider a system of $N$ qubits, for which the first $d$ qubits are in a pure state while the remaining $N-d$ qubits are in a fully mixed state. Let the state of the first $d$ qubits be pure, given by the partial density matrix $\ket{\rho_{[1,...,d]}}$ and that of all other qubits be mixed, given by $\frac{1}{2} I_{[j]}$ for $d \leq j \leq N$. The multi-qubit state, $\ket{\rho_m}$ is

\begin{equation}
    \ket{\rho_m}=\ket{\rho_{[1,...,d]}} \otimes \frac{1}{2^{N-d}} I_{[j]}^{\otimes (N-d)}.
\end{equation}

\noindent This means that the recurrence probability is

\begin{align}
    R_k=\bra{\rho_{m}} U^k \ket{\rho_{m}}\\
    =\frac{1}{2^{N-d}} \bra{\rho_{[1,...,d]}} U^k \ket{\rho_{[1,...,d]}}.
    \label{eq:subsystem}
\end{align}

\begin{figure}
	\includegraphics[width=0.97\columnwidth]{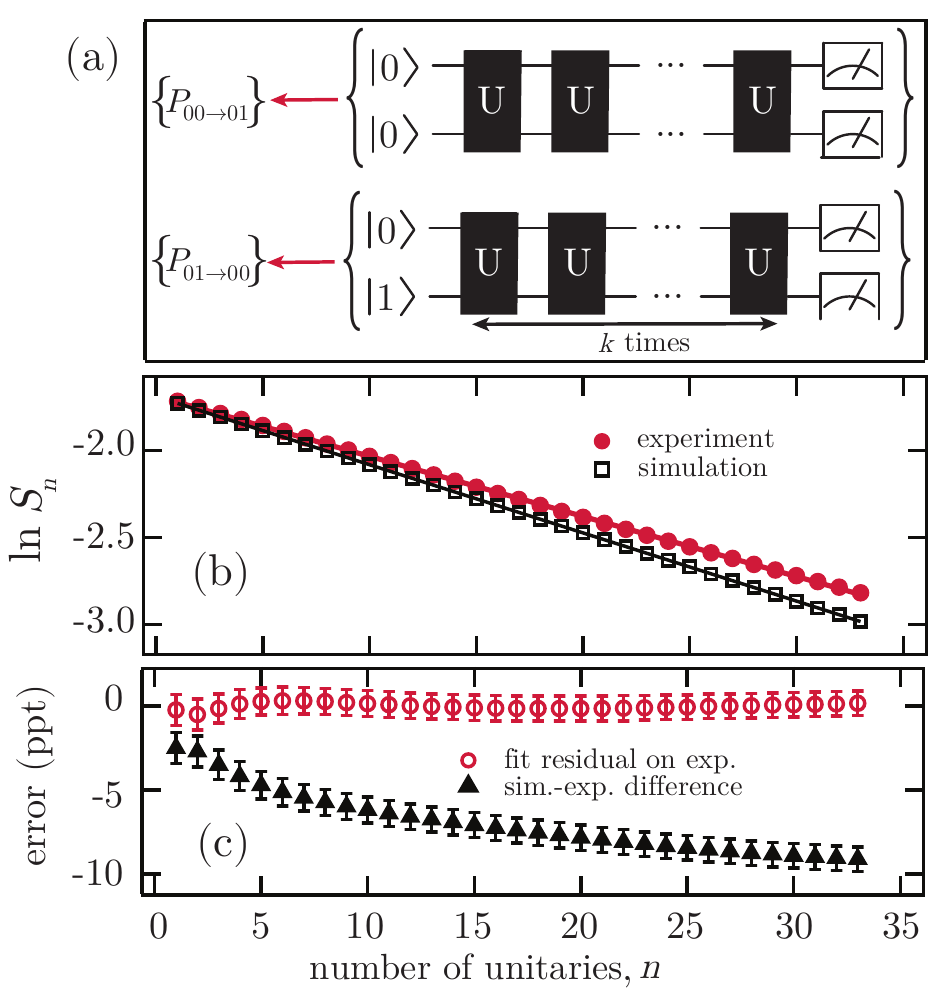}
    \caption{(a) The experimental setup for sub-system probes of unitarity: the first qubit as a pure state probe of the whole system. The second qubit is in an emulated fully mixed state, created by averaging the results of the two experiments shown. (b) Experimental measurement of $\ln S_n$ with the experimental results shown in closed red circles and the expected result from simulation in open black squares. Statistical error bars are smaller than the symbols and the solid lines are exponential fits. (c) Two types of errors on the value of $S_n$ given in parts per thousand: the difference between the experimental and simulated results is shown in closed black triangles and the residual from an exponential fit to $S_n$ is shown in open red circles.}
	\label{fig:Subsystem}
\end{figure}

\noindent A POE quantity based on this recurrence probability will exhibit exponential decay in the absence of POE-sensitive errors as all arguments used in Sec.~\ref{sec:ExpDecay} apply to Eq.~\ref{eq:subsystem} as well. In order to experimentally test this exponential decay, we emulate the creation of a mixed state and periodically evolve it in a system also containing a pure probe state. Figure~\ref{fig:Subsystem}a demonstrates the measurement scheme. The first qubit is a pure state probe which begins in $\ket{0}_{[1]}$. The second qubit is in a emulated mixed state with equal probabilities of being in $\ket{0}_{[2]}$ and $\ket{1}_{[2]}$. By repeating the experiment once with the second qubit in $\ket{0}_{[2]}$ and then again with the second qubit in $\ket{1}_{[2]}$, we can average the results to emulate the recurrence probability for a genuinely mixed state. As in previous sections, we drive this multiqubit state with the same periodic application of the unitary shown in Fig.~\ref{fig:LongEvolve}a. We then calculate $S_n$ for this recurrence probability and display the experimental results in red circles in Fig.~\ref{fig:Subsystem}b with errors plotted in Fig.~\ref{fig:Subsystem}c. As shown in the figure, the decay of $S_n$ is nearly exponential, indicating that in this range there are few POE-sensitive errors on either the probe qubit or the mixed-state qubit, similar to the pure state case in Fig.~\ref{fig:LongEvolve}. This shows that the protocol works on an entangled sub-system provided that the remainder of the system is in a mixed state. 

\subsection{Cross-state POE Constraints}

In this section, we choose two test states and show that the exponential decay law also applies to the probabilities of transitioning from one state to another after some time of periodic drive. Let the two states be $\ket{a}$ and $\ket{b}$. Recalling the expression in Eq.~\ref{eq:NegSum}, we measure $F^n$ as a transition probability from $\ket{a}$ to $\ket{b}$ which in Liouville space takes the following form.

\begin{equation}
\begin{gathered}
    \bra{b}F\ket{a}=\frac{1}{2^{2n}}\bra{a}F\ket{b}+\\
    \sum_{k=1}^{n} (-1)^k \binom{2n}{n-k} \bra{b} U^k \ket{a} + \\
    \sum_{k=1}^{n} (-1)^k \binom{2n}{n-k} \bra{b} U^{\dagger k} \ket{a}
    \label{eq:MeasTrans}
\end{gathered}
\end{equation}

\noindent Noting that the probability of making a transition from $\ket{a}$ to $\ket{b}$ after evolution $U^k$ is $P^{a \rightarrow b}_{k}\equiv \bra{b} U^k \ket{a}$ and similarly that $P^{b \rightarrow a}_{k}\equiv \bra{a} U^k \ket{b}=\bra{b} U^{-k} \ket{a}$, we find the following POE expression.

\begin{equation}
    T_n=C+\frac{1}{2^{2n}}\sum^{n}_{k=1}\binom{2n}{n-k} \frac{P^{a \rightarrow b}_{k}+P^{b \rightarrow a}_{k}}{2}
    \label{eq:tn}
\end{equation}

\noindent where $C=\frac{1}{2^{2n}}\binom{2n}{n}\braket{b}{a}$. We distinguish this quantity from $S_n$ in Eq.~\ref{eq:finalinequal} in that there is no inequality that can be derived from the positivity of $F^n$. However, $T_n$ does follow the same exponential decay as derived in Sec.~\ref{sec:ExpDecay} for $S_n$ as all arguments used in that section apply to $T_n$ as well.

\begin{figure}
	\includegraphics[width=0.97\columnwidth]{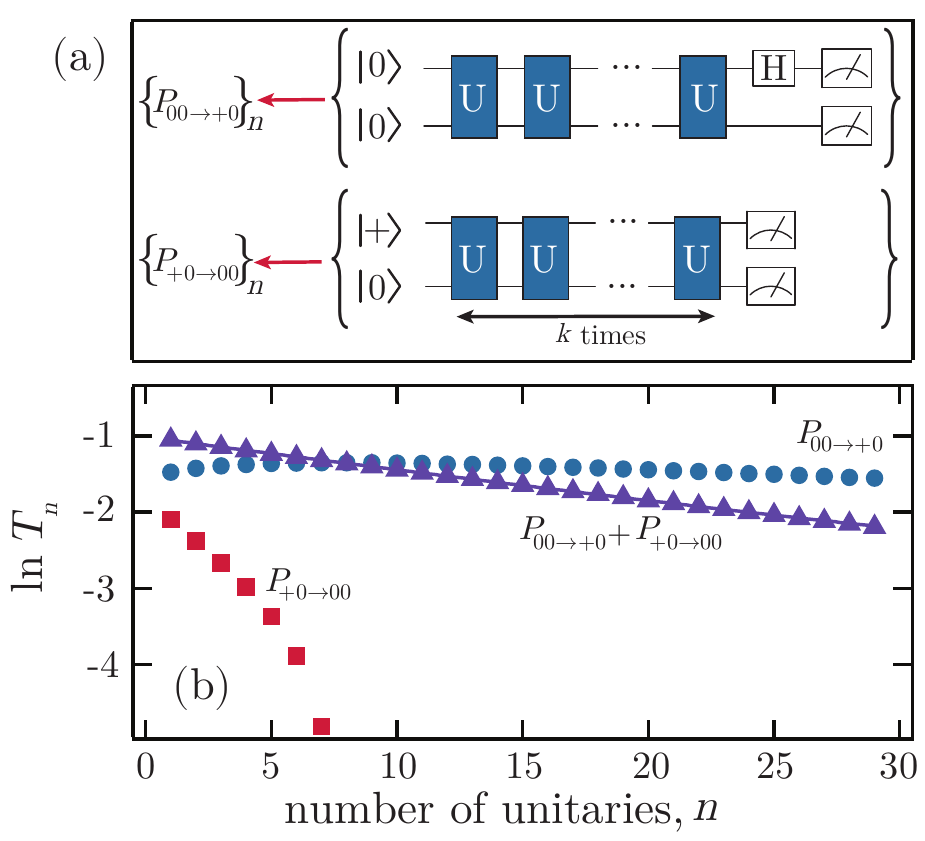}
    \caption{(a) The experimental sequence required for the cross-state POE measurement. The set $\{P_{k}^{+0\rightarrow 00}\}_n$ comes from one family of measurements in which the initial state is $\ket{+0}$ while the set $\{P_{k}^{00\rightarrow +0}\}_{n}$ comes from another family of measurements in which the initial state is $\ket{00}$ but the result is measured in the $\{\ket{+},\ket{-}\}$ basis on the first qubit using a Hadamard transformation. (b) Experimental results for the values of $\ln T_n$ (purple triangles) accompanied by the portion of the sum arising from $\{P_{k}^{+0\rightarrow 00}\}_n$ (red squares) and from $\{P_{k}^{00\rightarrow +0}\}_n$(blue circles). Note that while neither of the two contributions decay exponentially, their sum does as expected from Eq.~\ref{eq:tn}. The solid purple line is an exponential fit.}
	\label{fig:Transitions}
\end{figure}

We confirm the expected behavior of the cross-state POE constraint by measuring the two-way transition probabilities between the the states $\ket{0,0}$ and $\ket{0,+}$ under the same dynamics as shown in Fig.~\ref{fig:LongEvolve}. The experimental measurement scheme is shown in Fig.~\ref{fig:Transitions}a. Experimental results are shown in Fig.~\ref{fig:Transitions}b, along with an exponential fit showing that POE-sensitive errors are not visible on this scale. These results demonstrate that spectral analysis can be applied to more general POE constraints to probe the unitarity of quantum evolution. Whether specific POE constraints can yield insight beyond the character of the noise -- such as the location of the error or the degree of purity loss -- warrants further investigation~\cite{uzdin2021methods}.

\section{Conclusions\label{sec:Conclusions}}

In this paper we have derived a number of constraints on periodically driven quantum systems which distinguish some physical noise sources from others by being sensitive only to either incoherent errors or errors that break the periodicity of the drive to the system. We have also experimentally detected incoherent errors using a deviation from the expected exponential decay of the recurrence probability POE measurement. In addition to demonstrating the detection of intentionally added incoherent error, we have shown that similar exponential decay can be seen through spectral analysis of other POE constraints, including sub-systems that act as a probe for a whole system.

We note that because the number of unitary applications required to observe the signature of unitarity depends on the eigenvalue spectrum, rather than the system size directly, this signature will be experimentally feasible in some large systems. In particular, one can choose a tailored eigenvalue spectrum for diagnostic purposes which will exhibit the signature for a feasible number of measurements, making it a scalable diagnostic.

Finally, we note that there are important quantum algorithms which happen to be periodic, such as Hamiltonian simulation on either analog or digital devices. Hence, our method can be used to check for POE-sensitive errors as a secondary analysis to the results of such algorithms at no extra experimental overhead. 

\section*{Acknowledgements}

This work is supported by a collaboration between the US DOE and other Agencies. This material is based upon work supported by
the U.S. Department of Energy, Office of Science, National Quantum Information Science Research Centers, Quantum Systems Accelerator. Additional support is acknowledged from the National Science Foundation (QLCI grant OMA-2120757) and the Maryland—Army-Research-Lab Quantum Partnership (W911NF1920181).  A.M.G. was supported in part by a Joint Quantum Institute Postdoctoral Fellowship. RU is grateful for support from Israel Science Foundation (Grant No. 2556/20).

\appendix

\section{Construction of Inequalities \label{sec:coefderiv}}

In this section we detail the algebra required to derive Eq.~\ref{eq:NegSum}. Eq.~\ref{eq:FGton} can be rewritten as follows.

\begin{equation}
    F^n=\frac{1}{2^{2n}} \left( 2 - U - U^\dagger \right)^n
\end{equation}

\noindent Because $U$ is unitary, the above equation can be factored.

\begin{equation}
    F^n=\frac{1}{2^{2n}}\left( -U (1-U^\dagger)^2 \right)^n
\end{equation}

\noindent Expanding the product results in

\begin{align}
    F^n{}
    & =\frac{1}{2^{2n}} U^n (-1)^n \sum_{l=0}^{2n} \binom{2n}{l}\left( -U^{\dagger} \right)^{2n-l}\\
    & = \frac{1}{2^{2n}} \sum_{l=0}^{2n} \binom{2n}{l} (-1)^{n-l} \left( U^{\dagger} \right)^{n-l} .
\end{align}

\noindent Changing variables according to $k=n-l$, we find the following.

\begin{equation}
    F^n=\frac{1}{2^{2n}} \sum^{n}_{k=-n} \binom{2n}{n-k} (-1)^k U^k
\end{equation}

\noindent Finally, we split the sum and note that because $U$ is unitary, $U^{-k}=U^{\dagger k}$ and arrive at Eq.~\ref{eq:NegSum}.

\section{Limit to the Exponential Decay Law \label{sec:limit}}

In Eq.~\ref{eq:scaling} we showed that $\ln \left(S_n\right)$ decays linearly in $n$ for sufficiently large $n$. In this section, we specify how large $n$ must be. In order for $\lambda_{j_{\mathrm{max}}}$ to dominate the sum in Eq.~\ref{eq:EigSumRho}, we must satisfy

\begin{equation}
    \label{eq:linprecond}
    \lambda^{n}_{j_{\mathrm{max}}}|\braket{\rho_0}{j_{\mathrm{max}}} |^2 \gg \lambda^{n}_{j_{\mathrm{max,2}}}|\braket{\rho_0}{j_{\mathrm{max,2}}} |^2
\end{equation}

\noindent with $\lambda_{j_{\mathrm{max,2}}}$ being the second highest eigenvalue of $F$. Rearranging Eq.~\ref{eq:linprecond}, we find that the value of $n$ for which $S_n$ should decay exponentially is

\begin{equation}
    n\gg \frac{\ln\left(\frac{\left|\braket{\rho_0}{l_{j_{\mathrm{max,2}}}}\right|}{\left|\braket{\rho_0}{l_{j_{\mathrm{max}}}}\right|}\right)}{\ln\left(\frac{\lambda_{j_{\mathrm{max}}}}{\lambda_{j_{\mathrm{max,2}}}}\right)}.
\end{equation}

\noindent If the gap between $\lambda_{j_{\mathrm{max}}}$ and $\lambda_{j_{\mathrm{max,2}}}$ is small more cycles are needed for observing the exponential decay. Several eigenvalues could be clustered just below $\lambda_{j_{\mathrm{max}}}$ and well separated from the other eigenvalues. In this case, exponential decay of $S_n$ will be observed after the number of cycles is large enough to resolve the clustered eigenvalues. These limitations imply that it is possible that more measurements will be required as the size of the quantum system increases, depending on the spectrum of the unitary evolution. Finally, we note that arbitrarily increasing $n$ in order to check the exponential decay signature may be limited in practice by the variance of $S_{n}$ which does not decay exponentially.

\section{Robustness of Exponential Scaling Law to SPAM Errors \label{sec:SPAM}}

In this appendix we show that errors in state preparation and measurement do not change the exponential scaling of $S_n$. We will re-evaluate Eq.~\ref{eq:SnIneq} with SPAM errors included. For state preparation we model the preparation error as a sum of unitaries acting on the desired initial state prior to evolution with probability $p_{m}$~\cite{knill2005quantum}.

\begin{equation}
    \ket{\rho_0} \rightarrow \sum_m p_{m} U_{m} \ket{\rho_0}
\end{equation}

We model measurement error by assuming that the measured state is related to the actual state by a known detector matrix, $M$, which captures the probabilities of each computational basis state of being mistaken for each other possible basis state~\cite{nac20,nation2021scalable}. The actual state measurement is then:

\begin{equation}
    \bra{\rho_0} \rightarrow \bra{\rho_0}M.
\end{equation}

\noindent
Re-evaluating Eq.~\ref{eq:SnIneq} results in:

\begin{equation}
    S_n = \bra{\rho_0}F^{n}\ket{\rho_0} \rightarrow \sum_{m} p_{m} \bra{\rho_0} M F^{n} U_{m} \ket{\rho_0}
\end{equation}

\noindent Using the eigenbasis expression for $F^n$ gives the following form of $S_n$.

\begin{equation}
    S_n \rightarrow \sum_{m,j} p_{m} \bra{\rho_0} M \lambda_{j}^{n} \ket{j}\bra{j} U_{m} \ket{\rho_0}
\end{equation}

\noindent Again applying the approximation used in Eq.~\ref{eq:largen}, we arrive at the expression of $S_n$ in the case of sufficiently large $n$ with SPAM error included:

\begin{equation}
    S_n \rightarrow \lambda_{j_{\mathrm{max}}}^{n} \sum_{m} p_{m} \bra{\rho_0} M \ket{j_{\mathrm{max}}}\bra{j_{\mathrm{max}}} U_{m} \ket{\rho_0}.
\end{equation}

\noindent Applying the same reasoning as in Eq.~\ref{eq:scaling}, we see that $S_n \sim -n \ln{\frac{1}{\lambda_{j_{\mathrm{max}}}}}$. Therefore, SPAM errors may affect the constant offset on the value of $S_n$, but the exponential scaling is preserved.

\section{Proof that Sn is Monotonic \label{sec:monotonic}}

To prove that $S_n$ is monotonic under the POE assumptions, we consider the difference between two subsequent values of $S_n$ in the eigenbasis of $F$ as in Eq.~\ref{eq:EigSumRho}.

\begin{align}
    S_{n+1}-S_n=\sum^{N^2}_{j=1} \left( \lambda_{j}^{n+1} -\lambda_{j}^{n} \right) |\braket{\rho_0}{j} |^2 \\
    =\sum^{N^2}_{j=1} \lambda_{j}^{n}  \left( \lambda_{j} - 1 \right) |\braket{\rho_0}{j} |^2
\end{align}

We note that the last term in each element of the sum is trivially non-negative and that the first and second terms are also positive because $0 \leq \lambda_{j} \leq 1$ and conclude:

\begin{equation}
    S_{n+1}-S_n \leq 0
\end{equation}

Finally, we note that the same logic shows the second order derivative to be explicitly positive, such that the evolution of $S_n$ is concave under the POE assumptions.

\section{Experimental Details \label{sec:ExpDet}}

\begin{figure}
	\includegraphics[width=0.7\columnwidth]{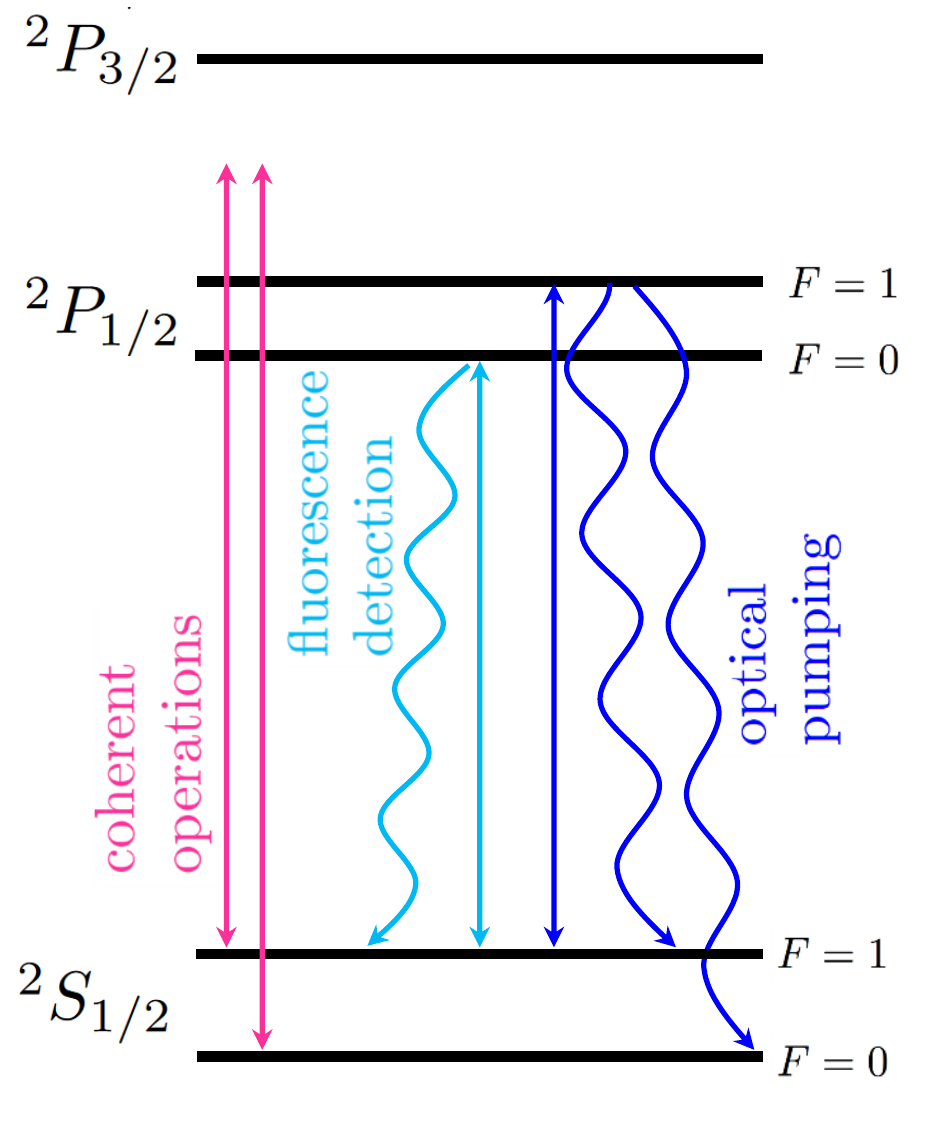}
    \caption{A simplified energy level diagram (not to scale) of $^{171}$Yb$^+$ showing the energy transitions addressed for state initialization, state detection, and coherent operations (quantum gates). Straight arrows indicate applied laser fields and wavy arrows indicate significant decay channels. The qubit states $\ket{0}$ and $\ket{1}$ are encoded in Zeeman sub-states no shown in the diagram.}
	\label{fig:Ybion}
\end{figure}

We collect experimental data using the TIQC first described in Ref.~\cite{deb16}, which is based on a chain of $^{171}$Yb$^+$ ions. The relevant electronic energy levels of $^{171}$Yb$^+$ are shown in Fig.~\ref{fig:Ybion} along with the laser fields required for qubit control operations. Excited state decay pathways are indicated with wavy arrows. The qubit states are encoded in two hyperfine-split ground electronic states of the $^{2}S_{1/2}$ manifold, $\ket{0}=\ket{F=0,m_{F}=0}$ and $\ket{1}=\ket{F=1,m_{F}=0}$. Qubits are initialized in $\ket{0}$ by optical pumping in the presence of a laser beam resonant with the $\ket{^{2}S_{1/2},F=1}\rightarrow \ket{^{2}P_{1/2},F=1}$ transition. State detection is performed through resonant fluorescence in the presence of a laser beam resonant with the $\ket{^{2}S_{1/2},F=1}\rightarrow \ket{^{2}P_{1/2},F=0}$ transition~\cite{olm07}. Coherent operations on the qubit states are performed with Raman laser beams far-off-resonantly driving both the $^{2}S_{1/2}\rightarrow ^{2}P_{1/2}$ and $^{2}S_{1/2}\rightarrow ^{2}P_{3/2}$ transitions.

For the purposes of adding tunable incoherent error, we apply the optical pumping beam typically used exclusively for state preparation. This beam is turned on at much lower than usual power to incoherently transfer population from $\ket{1}$ to $\ket{0}$ via the excited state $\ket{^{2}P_{1/2},F=1}$ at a chosen rate. In order to quantify the degree of the amplitude damping, we measure the lifetime of the $\ket{1}$ state in the presence of the pump beam alone.

The unitary evolution chosen to drive the TIQC is composed of two quantum gates as shown in Fig.~\ref{fig:LongEvolve}. These two gates are defined below.

\begin{align}
    XX\left(\theta\right)=e^{-i \frac{\theta}{2} \sigma_{x}^{j}\sigma_{x}^{k}}\\
    Y\left(\phi\right)=e^{-i\frac{\phi}{2}\sigma_{Y}^{j}},
\end{align}

\noindent where $j$ and $k$ label the qubits on which this gate acts and $\sigma$ is a Pauli spin operator along the specified axis. In this work, we chose $\theta=1.0$ rad and $\phi=2.4$ rad.

\end{document}